\documentclass[aps,prl,amsfonts,superscriptaddress,showpacs,twocolumn,byrevtex,nofootinbib]{revtex4}

\usepackage{epsfig}
\usepackage{amsmath}
\usepackage{amssymb}
\usepackage{amsfonts}
\usepackage{color}

\allowdisplaybreaks     

\newcommand{\PGRcomm}[1]{{\textcolor{black}{ #1}}}
\newcommand{\EScomm}[1]{{\textcolor{black}{ #1}}}

\newcommand{\JMcommNEW}[1]{{\textcolor{black}{ #1}}}

\begin{document}

\title{Time-dependent Generalized SIC-OEP formalism and Generalized SIC-Slater approximation.}

\author{J. Messud}
\affiliation{
Universit\'e de Toulouse; UPS; Laboratoire de Physique
  Th\'eorique (IRSAMC); F-31062 Toulouse, France
}
\affiliation{
CNRS; LPT (IRSAMC); F-31062 Toulouse, France
}

\author{P. M. Dinh}
\affiliation{
Universit\'e de Toulouse; UPS; Laboratoire de Physique
  Th\'eorique (IRSAMC); F-31062 Toulouse, France
}
\affiliation{
CNRS; LPT (IRSAMC); F-31062 Toulouse, France
}

\author{P.-G. Reinhard}
\affiliation{Institut f\"ur Theoretische 
  Physik,
  Universit{\"a}t Erlangen, D-91058 Erlangen,
  Germany}

\author{Eric Suraud}
\affiliation{
Universit\'e de Toulouse; UPS; Laboratoire de Physique
  Th\'eorique (IRSAMC); F-31062 Toulouse, France
}
\affiliation{
CNRS; LPT (IRSAMC); F-31062 Toulouse, France
}


\date{\today} 
\begin{abstract}
We propose a simplification \EScomm{of the full ``2 sets'' Time dependent Self Interaction 
Correction (TD-SIC) method, }
applying the Optimized Effective Potential (OEP)
method. The new resulting scheme is called time-dependent ``Generalized SIC-OEP''.
A straightforward approximation, using the spatial localization of one set of orbitals, leads to the ``Generalized SIC-Slater'' formalism.
We show that it represents a great improvement compared to the traditional SIC-Slater/KLI formalisms.
\end{abstract}

\maketitle

Density Functional Theory
(DFT)~\cite{Koh99,DFT}
has become a standard theoretical tool for the
description of electronic properties in a variety of  physical and chemical
systems, especially for sizable numbers of electrons. The
extension of DFT to Time-Dependent (TD) situations~\cite{gross} is a
more recent achievement which also  motivates numerous  
formal and practical investigations \cite{Mar06}.
TDDFT contitutes to date one of the few, well founded
theories, for describing dynamical scenarios in complex systems.
{From the point of view of applications TDDFT
requires simple approximations, 
the simplest one being the} \JMcommNEW{Adiabatic}
Local Density Approximation (\PGRcomm{A}LDA), 
which proved very useful in calculations of structure and low-amplitude excitations
(optical response, direct one-photon processes) \cite{Koh99}.
{But self-interaction error plagues ALDA, which 
in particular spoils the dynamical description  of
excitations involving ionization, especially when close to 
threshold. A correct treatment requires a self-interaction correction (SIC), such 
the one proposed in \cite{Per81}. }
This SIC  has been used since then at various levels of refinement for structure
calculations in atomic, molecular, cluster and solid state physics,
see e.g. \cite{Ped84,Goe97,Polo,Vydrov}. 
{But the SIC scheme leads to an orbital dependent mean field, 
which causes several formal and technical difficulties
\cite{Din08}, becoming all the more cumbersome in the
TD case.}
{Applications of SIC in TD situations are thus mostly done in
approximate manner, linearized \cite{Pac92}, } using
averaged-density SIC \cite{Leg02a}, or relying on various approximations of
TD Optimized Effective Potential (TDOEP) scheme \cite{Ull95a} {(the latter
allows to deal with a common local potential).} 
The most popular {approximate OEP scheme} is the Krieger-Li-Iafrate (KLI)
approach \cite{Kue08,Kri92,Ton} and, in a further step of simplification,
the Slater approximation \cite{Sha53a}. {But both suffer from inconsistencies,}
such as violation of zero force theorem and energy conservation
\cite{Mun07a}, even if some applications may be more forgiving~\cite{Wij08}.

{We have recently} generalized the considerations of 
{stationary SIC} \cite{Ped84} to
the {time dependent} case \cite{Mes08-1} 
and showed that the time propagation of the ``exact'' SIC scheme
{can be performed thanks to  a double basis set formulation
(both sets being connected by a unitary transformation building the
same total density).}
The resulting scheme satisfies all key formal properties.
But the remaining potential becomes non-local, which substantially
slows down numerical calculations.
{In the stationary case, a} {SIC-OEP scheme relying on
this two sets technique can be performed leading to ``Generalized
SIC-OEP'' either in full formulation \cite{Kor08-2} or in the
``Generalized SIC-Slater'' (GSlat) approximation \cite{Mes08-2} which
emerges naturally because of the localization of one set of orbitals
\cite{Mes08-2}. GSlat is numerically efficient and represents a great
formal and practical improvement over traditional stationary
SIC-Slater/KLI \cite{Mes08-2,Mes09-2}. It remains to extend Generalized SIC-OEP to the
time domain and consider simple efficient approximations thereof, such
as possibly the time-dependent GSlat (TDGSlat). {It} is the
aim of this paper to address both aspects.}


%
The starting point is the
SIC energy functional for electrons
(all sums run over occupied states only ; we omit the ($\mathbf{r},t$) 
dependencies when it is not misleading and we note
$\rho=\sum_\alpha |\psi_\alpha|^2$)
\begin{eqnarray}
  &&
  E_\mathrm{SIC} =
\label{eq:fsicen}\\
  &&
  \sum_\alpha (\psi_\alpha|\frac{\mathbf{p}^2}{2m}|\psi_\alpha)
  \!+\!
  E_\mathrm{ext}[\rho]
  \!+\!
  E_\mathrm{ALDA}[\rho]
  \!-\!\sum_\alpha E_\mathrm{ALDA}[|\psi_{\alpha}|^2].
\nonumber
\end{eqnarray}
{The first term is the {non-interacting} kinetic energy,
$E_\mathrm{ext}[\rho]=\int d\mathbf{r}\rho v_{\rm ext}$ where $v_{\rm ext}$
stands for the interaction with the ionic background and any other
local possibly time-dependent external field, and}
$E_\mathrm{ALDA}[\rho]$ is a standard ALDA energy-density functional
The last term corresponds to the SIC. 
\PGRcomm{We mention in passing that the SIC, and with it all our
following development, does also apply to more general funtionals as,
e.g., the Generalized Gradient Approximation (GGA) \cite{Per96a}.}
The TDSIC equations result from a variational principle
\PGRcomm{applied to the action integral}
\JMcommNEW{(here we use the standard action functional.
Its causality and boundary conditions problems are discussed in
\cite{Gro95,Lee98}). 
}
\PGRcomm{Furthermore, we
explicitly include} the orthonormality constraint with Lagrange
multipliers.
\PGRcomm{This yields}
\begin{eqnarray}
 && \delta \left( A_{\rm SIC} - \int_{-\infty}^{t_1} \textrm dt
  \sum_{\beta,\gamma}^{}(\psi_{\beta}|\psi_{\gamma})
  \lambda_{\gamma\beta} \right) = 0 \quad ,
  \nonumber\\
 && A_{\rm SIC}
 =
 \int_{-\infty}^{t_1} \textrm dt \Big(
  E_\mathrm{SIC}
  -
 \sum_{\alpha}(\psi_\alpha|\mathrm{i}\hbar \partial_t|\psi_{\alpha})
 \Big) \quad ,
\label{eq:varprinconstr}
\end{eqnarray}
leading to one-body equations that we recast as~\cite{Mes08-1}
\begin{eqnarray}
  && 
  \mathrm{i} \hbar \partial_t |\varphi_i) = \hat{h}_\mathrm{SIC}|\varphi_i)
\label{eq:td-diag} \\
  && \forall t : \hspace{1mm} 0
  =
  (\psi_\beta|U_{\rm ALDA} \left[ |\psi_\beta|^2 \right]-U_{\rm ALDA} \left[ |\psi_\alpha|^2 \right]|\psi_\alpha)
\label{eq:symcond2}
\end{eqnarray}
where $U_{\rm ALDA} \left[ \rho_\alpha \right]= (\delta E_\mathrm{ALDA}[\tilde\rho]/\delta
\tilde\rho)_{|\tilde\rho=\rho_\alpha}$ and 
\begin{eqnarray}
  &&\hat{h}_\mathrm{SIC}
  =
  \hat{h}_\mathrm{ALDA}
  -
  \sum_\alpha U_{\rm ALDA} \left[ |\psi_\alpha|^2 \right]|\psi_\alpha)(\psi_\alpha|
\label{eq:hsic}\\
  &&\hat{h}_\mathrm{ALDA}
  =
  \frac{\hat{\mathbf{p}}^2}{2m}
  +
  v_{\rm ext}
  +
  U_{\rm ALDA} \left[ \rho \right].\nonumber 
%
\nonumber
\end{eqnarray}
%

Note that $\hat{h}_{\rm SIC}$ is applied to the
\textit{diagonal orbitals} $|\varphi_i)$ {in} Eq. (\ref{eq:td-diag}), but
is calculated with the \textit{localized orbitals}
$|\psi_\alpha)$ {in} Eq. (\ref{eq:hsic}).
The $|\varphi_i)$'s are related to the $|\psi_\alpha)$'s by a unitary
transformation 
\begin{eqnarray}
\label{eq:unitrans}
  \psi_\alpha  =  \sum_{i} \varphi_i \, u_{i\alpha}
  \hspace{2mm};\hspace{2mm}
  \sum_{i} u_{i\alpha} u^*_{i\beta}=\delta_{\alpha\beta}\quad.
\end{eqnarray}
\PGRcomm{The wave functions $\varphi_i$ and the
transformation coefficients $u_{i\alpha}$ are varied
independently where variation with respect to  $\varphi_i$
yields Eq.~(\ref{eq:td-diag}) and variation
of the  $u_{i\alpha}$ then leads to}
the symmetry condition (\ref{eq:symcond2}) \cite{Mes08-1,Mes09}.
{The symmetry condition guarantees conservation of
orthonormality and enforces localization of the states
$|\psi_\alpha)$} \cite{Mes09}. It thus plays a key role in this
formalism.
%
%
Although Eq.~(\ref{eq:td-diag}) has proven to be {tractable}
numerically \cite{Mes08-1,Mes09}, the corresponding Hamiltonian is
non-local, see Eq.~(\ref{eq:hsic}), which implies a very high
numerical cost.  

To {develop a simplified scheme while maintaining locality,
we apply this double set formulation to the TDOEP scheme.}
%
We start from the action integral~(\ref{eq:varprinconstr}). The
previous considerations on the ``exact'' SIC energy showed that it
should be expressed with the localized $\psi_\alpha$, which satisfy
the symmetry condition (\ref{eq:symcond2}). We stationarize the action, 
imposing that the diagonal $\varphi_i$ (linked by a unitary
transformation to the $\psi_\alpha$) satisfy a Kohn-Sham like equation
{with a local and state-independent potential $V_0$} 
{(although OEP orbitals are not exactly the same as the SIC ones,
we employ 
the same notations to {keep the presentation compact})}%
%
\begin{eqnarray}
  \left[\hat{h}_\mathrm{ALDA}-V_0(\mathbf{r},t)\right] \varphi_i({\bf r},t)
  =
  \mathrm{i}\hbar\partial_t \varphi_i({\bf r},t)
  \quad.
\label{eq:eigen3}
\end{eqnarray}
The orthonormality constraint is implicitely contained in (\ref{eq:eigen3}).
The result will thus be considered as a local approximation of the SIC Hamiltonian (\ref{eq:hsic}).
The optimized effective potential $V_0(\mathbf{r})$ is found by the
variation $\delta A_{\rm SIC} / \delta V_0(\mathbf{r},t)=0$
\PGRcomm{which employs variations with respect to the
$\varphi_i$ through the chain rule for functional derivatives.}
The independent variation of the coefficients $u_{i\alpha}$
in the transformation (\ref{eq:unitrans}) remains as before.
We obtain
\cite{Ull95a}
%
\begin{eqnarray}
&& \sum_i \int_{-\infty}^{t_1} \textrm dt' \int \textrm d\mathbf{r'}
  \Big( V_0(\mathbf{r'},t')-v_i^*(\mathbf{r'},t') \Big) \nonumber\\
&& \hspace{5mm}\times K_i(\mathbf{r},t ; \mathbf{r'},t')\varphi_i^*(\mathbf{r'},t') \varphi_i(\mathbf{r},t) +c.c. =0
\label{eq:tdOEP2}\\
&& K_i(\mathbf{r},t ; \mathbf{r'},t')=- \mathrm{i} \sum_{j=1,j\ne
  i}^{+\infty}\varphi_j^*(\mathbf{r},t)\varphi_j(\mathbf{r'},t')\theta(t-t') 
\\
&& v_i(\mathbf{r},t)= \frac{1}{\varphi_i (\mathbf{r},t)}
\frac{\delta}{\delta \varphi_i^* (\mathbf{r},t)} \int_{-\infty}^{t_1}
\textrm dt' \sum_\alpha E_{\rm ALDA}[|\psi_\alpha|^2](t')
\nonumber\\ 
&& \hspace{0.6cm} = \frac{1}{\varphi_i(\mathbf{r},t)} \sum_{\alpha}
\upsilon_{i \alpha}^{*}(t) U_{\rm ALDA} \left[ |\psi_\alpha|^2
  \right](\mathbf{r},t) \psi_\alpha(\mathbf{r},t)
\label{eq:v_i-td}
\end{eqnarray}
to be fulfilled together with the symmetry condition
 (\ref{eq:symcond2}) \PGRcomm{which again results
from variation of the $u_{i\alpha}$}.
This is the TD ``Generalized SIC-OEP'' formalism.
{The new feature in this double-set TDOEP appears in the $v_i$
in (\ref{eq:v_i-td}) which now employs the localized
$\psi_\alpha$, in accordance with exact
TDSIC.}  The previous applications of TDOEP to SIC as found in
\cite{Ton} used the action ~(\ref{eq:varprinconstr}) written in terms of the diagonal
$\varphi_i$, which leads to well known pathologies, as the
incapability of the traditional SIC-Slater/KLI approximations to
reproduce the spatial localization.
Eq. (\ref{eq:tdOEP2}) can be rewritten equivalently as
\begin{eqnarray}
V_0 = \Re e \{ V_{\rm S}+V_{\rm K}+V_{\rm C} \} - \Im m \{ V_{\rm TD1}+V_{\rm TD2} \} \nonumber
\label{eq:TD-OEP_gen}
\end{eqnarray}
where $V_{\rm S}$, $V_{\rm K}$, $V_{\rm C}$ are defined as 
\begin{eqnarray}
V_{\rm S} &=& \sum_i \frac{|\varphi_i|^2}{\rho}v_i ,\\
V_{\rm K} &=& \sum_i
\frac{|\varphi_i|^2}{\rho}(\varphi_i|V_0-v_i|\varphi_i) ,\\
V_{\rm C} &=& \frac{1}{2}\sum_i \frac{\mathbf{\nabla}.(p_i
  \mathbf{\nabla}|\varphi_i|^2)}{\rho},
\label{eq:pot_OEP}\\
p_i(\mathbf{r},t)&=& \frac{1}{\varphi_i^*(\mathbf{r},t)}
\int_{-\infty}^{t_1} \textrm dt' \int \textrm d\mathbf{r'} 
\Big( V_0(\mathbf{r'},t')-v_i^*(\mathbf{r'},t') \Big) 
\nonumber\\
&& \hspace{2cm} \times\varphi_i^*(\mathbf{r'},t') K_i(\mathbf{r},t ;
\mathbf{r'},t'). 
\label{eq:p_i-td}
\end{eqnarray}
To those potentials, which also appear in the stationary case, one has
to add purely time-dependent contributions
 
\begin{eqnarray}
V_{\rm TD1} &=& \frac{1}{\rho}\sum_i
\frac{\mathbf{\nabla}^2|\varphi_i|^2}{4} \int_{-\infty}^{t} \textrm
dt' (\varphi_i(t')|v_i(t')|\varphi_i(t')) \nonumber\\
V_{\rm TD2} &=& \frac{1}{\rho}\sum_i \left(   |\varphi_i|^2
\frac{\partial p_i}{\partial t} + \mathbf{J}_i.\mathbf{\nabla} p_i
\right)
\label{eq:pot_OEP1-td}
\end{eqnarray}
where $\mathbf J_i=\frac{\hbar}{2 \mathrm{i} m}( \varphi_i^* \nabla\varphi_i
- \varphi_i \nabla\varphi_i ^*)$ is the current density.
Some straightforward manipulation with the unitary transformation
(\ref{eq:unitrans}) allow to rewrite
\begin{eqnarray}
&& 
V_{\rm S} = \sum_\alpha \frac{|\psi_\alpha|^2}{\rho} U_{\rm
    ALDA}[|\psi_\alpha|^2] \hspace{3mm} \in \Re e ,
\nonumber\\
&& 
V_{\rm K} =\frac{1}{\rho} \sum_{\alpha,\beta} \Big( \sum_i
|\varphi_i|^2 \upsilon_{i\alpha}^{*}\upsilon_{i\beta} \Big) ({
  \psi_\beta}|V_0-U_{\rm ALDA} [|\psi_\alpha|^2]|\psi_\alpha) ,
\nonumber\\
&&
p_i(\mathbf{r},t)= \frac{1}{\varphi_i^*(\mathbf{r},t)}
\sum_\alpha\ \upsilon_{i\alpha}(t) \int_{-\infty}^{t_1} \textrm dt'
\int \textrm d\mathbf{r'} \psi_\alpha^*(\mathbf{r'},t')
\nonumber\\
&& \hspace{5mm}\times 
\Big( V_0(\mathbf{r'},t')-U_{\rm ALDA}
    [|\psi_\alpha|^2](\mathbf{r'},t') \Big)
     K_i(\mathbf{r},t ; \mathbf{r'},t') .
\label{eq:pot_OEP2-td}
\end{eqnarray}
Note that the ``Generalized SIC-KLI'' approximation $V_{\rm K}$ has not the form 
that might be intuitively expected \cite{Kor08-2}. 
Moreover, as $(\varphi_i|v_i|\varphi_i)=(\varphi_i|v_i|\varphi_i)^*$, we have, with
(\ref{eq:pot_OEP1-td}), $\Im m \{ V_{\rm TD1} \}=0$.

{It turns out that} the full TD ``Generalized SIC-OEP''
scheme is very costly numerically, even more than the ``exact'' TDSIC
formalism because
\JMcommNEW{time integrals (memory effects) appear explicitly in the definition of $V_0$.}
Thus we propose a {strong simplification} which emerges naturally
from the fact that the $\psi_\alpha$ remain spatially very localized
\cite{Ped84,Mes09}. This means that, at {all time, at} a given
$\mathbf r$, one single $\psi_\alpha$ mostly dominates the other wave
functions $\psi_{\beta\ne\alpha}$ and thus
\begin{eqnarray}
\sum_\beta
\frac{|\psi_\beta|^2}{\rho} U_{\rm  ALDA}
     [|\psi_\beta|^2]\psi_\alpha
\approx U_{\rm ALDA}[|\psi_\alpha|^2]\psi_\alpha.
\label{eq:loc}
\end{eqnarray}
One finds that the Slater contribution $V_{\rm S}$ is 
dominating, although the other terms are generally non negligible.
But in the TD ``Generalized SIC-OEP'', if we approximate
$
  V_0
  \approx
  V_{\rm S} = \sum_\alpha \frac{|\psi_\alpha|^2}{\rho}U_{\rm ALDA}[|\psi_\alpha|^2]
\label{eq:h_oep-td}
$,
the localization (i.e. Eq.~(\ref{eq:loc})) amounts to have almost vanishing
$V_{\rm K}$ and $p_i$, see Eq.~(\ref{eq:pot_OEP2-td}), and thus almost
vanishing $V_{\rm C}$ and $V_{\rm TD2}$, see Eqs.~(\ref{eq:pot_OEP}) and
(\ref{eq:pot_OEP1-td}).
Thus, the TDOEP result naturally reduces to
\begin{eqnarray}
  V_0
  &\simeq&
  \sum_\alpha \frac{|\psi_\alpha|^2}{\rho}U_{\rm ALDA}[|\psi_\alpha|^2]
  \quad.
\label{eq:Slater3}
\end{eqnarray}
%
Eq.~(\ref{eq:eigen3}) generates the set $\varphi_i$ of diagonal states,
which can be interpreted (to first order) as single electron orbitals,
while the unitary transformation (\ref{eq:unitrans}) serves to accommodate the symmetry condition
(\ref{eq:symcond2}) which, in turn, defines the localized orbitals
$\psi_\alpha$ entering the potential $V_0$ as given in Eq.~(\ref{eq:Slater3}).
Note that this equation has the form of a Slater
approximation \cite{Kri90a} but is constructed from the localized $\psi_\alpha$ and applied 
to the $\varphi_i$. This is the time-dependent extension of the
``Generalized SIC-Slater'' scheme (TDGSlat). 
We showed in \cite{Mes08-2} that its stationary counterpart
solves many {problems} encountered with traditional
SIC-Slater/KLI methods. 

\begin{figure}[htbp]
\begin{center}
\includegraphics[width=7.5cm,height=4.5cm,angle=0]{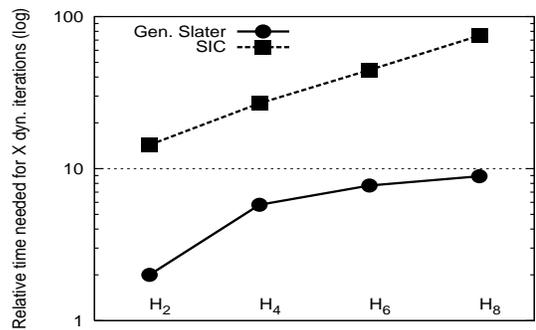}
\caption{Gained time in TDGSlat compared to TDSIC (in unit of time for ALDA propagation 
{in $H_n$:
$t(\mathrm{H}_n^\mathrm{(method)})/t(\mathrm{H}_n^\mathrm{(ALDA)})$, and logarithmic scale}); calculations on H chains as indicated.
\label{fig:time}}
\end{center}
\end{figure}
%
%
We first check the performance of TDGSlat as compared to full
TDSIC in Fig. \ref{fig:time}
Times are shown relative to the time needed to propagate the cheapest
solution (ALDA) for Hydrogen chains on a given physical time
interval (mind the logarithmic scale). The gain as compared to TDSIC
is dramatic (typically of an order of magnitude) for a cost an order
of magnitude larger than ALDA, for H$_8$.

%
\begin{figure}[htbp]
\begin{center}
\includegraphics[width=7.5cm,height=6.5cm,angle=0]{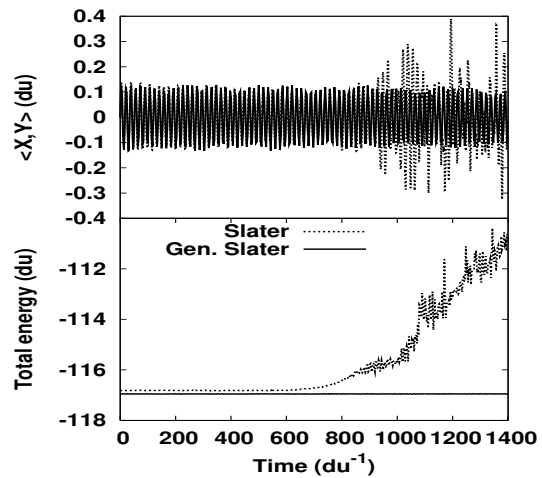}
\caption{Total energy and dipole moment (along perturbation) in a quantum dot 
as a function of time in the linear domain. See
text for details. 
\label{fig:qdot}}
\end{center}
\end{figure}

%
%
%
 
%
\begin{figure}[htbp]
\begin{center}
\includegraphics[width=7.5cm,height=6.5cm,angle=0]{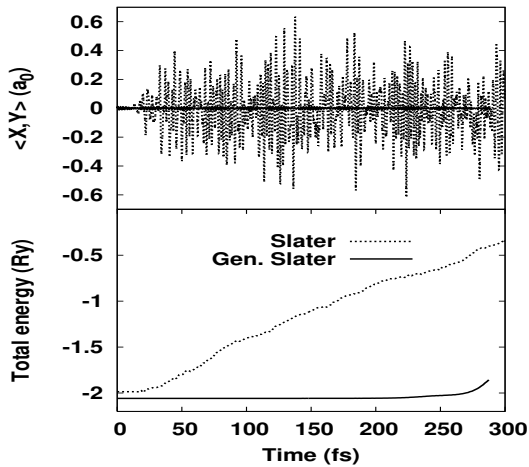}
\caption{Same as Figure  \ref{fig:qdot} but for a Na$_5$ cluster.
\label{fig:na5}}
\end{center}
\end{figure}

{It is well known that standard (one set) 
SIC-Slater and KLI approximations suffer from 
inconsistencies in the time domain.} They can strongly
violate energy conservation and zero force theorem (ZFT)
\cite{Mun07a}, while {neither full TDSIC \cite{Mes09} nor full
``Generalized SIC-OEP'' do suffer from these defects, but at the price of 
very heavy numerical cost.  
Strictly speaking, at TDGSlat level one has}
(for a non-explicitly TD external potential)
\begin{eqnarray}
&&\partial_t E_\mathrm{SIC}
=\frac{\hbar}{m}
  \sum_\alpha \Im m \int \textrm d\mathbf{r} 
\nonumber\\
&&\hspace{5mm}\times\{U_{\rm ALDA}[|\psi_\alpha|^2]-\sum_\beta
\frac{|\psi_\beta|^2}{\rho}U_{\rm
  ALDA}[|\psi_\beta|^2]\}\psi_\alpha\Delta\psi_\alpha^* ,
\nonumber\\
&&\partial_t \sum_i(\varphi_i|\mathbf{p}|\varphi_i)=\int \textrm
d\mathbf{r} \, v_{ext} \nabla\rho + \sum_\alpha \Re e \int \textrm d\mathbf{r}
\nonumber\\
&& \hspace{5mm}\times\{U_{\rm ALDA}[|\psi_\alpha|^2]- \sum_\beta
\frac{|\psi_\beta|^2}{\rho} U_{\rm
  ALDA}[|\psi_\beta|^2]\}\psi_\alpha\Delta\psi_\alpha^* .
\nonumber
\end{eqnarray}
{which in principle leads to energy and ZFT violation as
well. However,} the localization of the $\psi_\alpha$, {which
implies the approximate relation (\ref{eq:loc}),} lets us hope that
{ TDGSlat produces ``less'' violated energy conservation
and zero-force theorem, where ``less'' means that the conservation
laws remain stable for a longer time span than in simple
 SIC-Slater/KLI. This alone could be progress for many practical
purposes.}
It is thus a key issue to explore how TDGSlat practically performs in
the time domain.  {This has been done in a variety of systems
and we illustrate our results in Figures \ref{fig:qdot} and
\ref{fig:na5} on two typical cases: a quantum dot with 6 electrons,
in the spirit of \cite{Pi04}, and
a small metallic cluster. Calculations have been done in full 3D using
the same numerical methods as in \cite{Mes09}. In both examples we
plot the {time evolution of the} total energy and the dipole
moment after a small boost of the initial (ground state) electronic
distributions which simulates a {very short laser pulse}, and
still allows to check energy conservation in time {because the
excitation field is switched off during propagation}.  We take care
of considering sufficiently moderate perturbations to remain in the
linear domain (excitation energy a few percent of the typical
electronic level spacing) and we follow the dynamics over "long"
times, typically over 100 eigenperiods of the system, as seen from the
dipole oscillations. We restrict the calculations to TDGSlat and
SIC-Slater. {Full} SIC calculations on such long times are
prohibitively costly.  We have checked on shorter times that they
deliver perfect energy and ZFT conservation.}

{Both figures deliver the same message and we thus discuss
them {together}. While both full SIC and TDGSlat dynamics
remain remarkably stable in time, standard SIC-Slater exhibits a
sizeable violation of energy and ZFT
(seen through the unstability of the dipolar moment \cite{Mun07a}).
In the case of energy one
observes a strong drift in time while for the dipole moment
oscillations become extremely large (much larger than the original
ones). When pursuing the TDGSlat over even longer times one observes a
small energy violation, as expected on formal grounds.  Still the
effect is rather small and much delayed as compared to standard
SIC-Slater. And the most important point is that the drift occurs on
sufficiently long times so that relevant physics can be studied for
shorter times. One can thus conclude that dynamics in the linear
domain is controlled in TDGSlat, remaining very close to full TDSIC
over long times.}

%
To summarize, we started from the two basis set formulation of
TDSIC and applied the TDOEP formalism to recover locality, resulting
to the time-dependent ``Generalized SIC-OEP'' formalism. 
As it is very costly numerically, we looked for a much less costly
relevant approximation, which naturally comes from the localized
character of the $\psi_\alpha$ set and is called time-dependent
``Generalized SIC-Slater''.
By virtue of the double-set technique, it has a wider range of
applicability than traditional SIC-Slater/KLI approximations.
In particular, we checked formally and numerically on various systems (organic, metal, quantum dot)
that TDGSlat will satisfy much better energy and ZFT conservation.

This work was supported,
by  
%
Agence Nationale de la Recherche (ANR-06-BLAN-0319-02), 
the Deutsche Forschungsgemeinschaft (RE 322/10-1),
and the  Humboldt foundation.

\end{document}